\documentstyle[psfig]{elsart}

\newcommand{\bd}[1]{ \mbox{\boldmath $#1$}  }
\begin{document}

\begin{frontmatter}
\title{The $^{10}$Li spectrum and the $^{11}$Li properties }

\author{E. Garrido}
\address{Instituto de Estructura de la Materia, CSIC, Serrano 123, E-28006
Madrid, Spain}
\author{D.V.~Fedorov \and A.S.~Jensen}
\address{Institute of Physics and Astronomy,
University of Aarhus, DK-8000 Aarhus C, Denmark}
\date{\today}

\maketitle

\begin{abstract}
The neutron--$^9$Li interaction and the corresponding low-energy
$^{10}$Li spectrum are decisive for the properties of $^{11}$Li
described as a three-body system (n+n+$^{9}$Li). We compute structure
and breakup properties of $^{11}$Li as function of this interaction.
The hyperfine structure due to the spin $3/2$ of both $^{9}$Li and
$^{11}$Li is needed and treated with special care.  We use the
hyperspherical adiabatic expansion of the Faddeev equations for the
structure and the participant--spectator model for the breakup
reactions of $^{11}$Li. We use established experimental constraints of
both $^{11}$Li (binding energy, size and differential breakup cross
sections) and of $^{10}$Li (a virtual $s$--state below $50$~ keV and
a $p$--resonance around 0.54 MeV)\footnote{
Bound and virtual states have positive and negative imaginary wave numbers, 
respectively (zero real parts). Thus the virtual $s$-state
energies should strictly speaking be negative, but we shall here and
throughout the paper omit the minus sign and use the absolute value.
}. 
Another $p$--resonance must then be
present below 0.54 MeV and another $s$--level must be present between
about 0.5 MeV and 1 MeV depending on unknown spin assignments.  All
established facts are in agreement with our predictions obtained
within the same consistent model.
\end{abstract}

\end{frontmatter}

\par\leavevmode\hbox {\it PACS:\ }  21.45.+v, 25.10.+s, 25.60.-t,  25.60.Gc

\section{Introduction and motivation}

The $^{11}$Li nucleus is a prominent example of the so-called
Borromean nuclei which can be described as a three--body system 
($^{9}$Li plus two neutrons) where none of the internal two--body subsystems 
is bound \cite{zhu93}. Few-body models and techniques have been very successful
\cite{nie01} but they must necessarily rely heavily on the underlying
two-body interactions. Phenomenological interaction parameters for the
neutron-$^{9}$Li system have usually been obtained by constraints of
the properties of $^{11}$Li. However, comparison between measurements
and these theories suffered from the fact that essentially all
theoretical investigations assumed spin zero for both $^{9}$Li and
$^{11}$Li.  This is very unsatisfactory and also rather surprising
especially considering the huge amounts of work invested in
$^{11}$Li. Substantially more complicated models are required but the
available techniques are also very developed. The new experimental
results have prompted us to reinvestigate the present
$^{10}$Li-$^{11}$Li connection.

\subsection{The two-body system}

The experimental data about the properties of $^{10}$Li, and therefore
also of $^{11}$Li, still remain incomplete.  The quantum numbers of
the $^{10}$Li ground state are not firmly established.  The first
attempt to measure the $^{10}$Li spectrum was made by Wilcox et
al. \cite{wil75} placing the ground state of the unbound nucleus
$^{10}$Li at 0.80$\pm$0.25 MeV.  More recent experiments \cite{ame90}
concluded the existence of a very low-lying state in $^{10}$Li,
although its quantum numbers were not determined.  In \cite{boh93} the
authors identified the 1$^+$--state as the most probable $^{10}$Li
ground state. This parity inversion was already discussed much earlier
for neighboring nuclei in \cite{tal60} and for $^{10}$Li in
\cite{bar77}.

A probable $s$--state as ground state of $^{10}$Li was experimentally
suggested in \cite{kry93}. In \cite{you94} a $p$--resonance was
observed at an energy of 538$\pm$62 keV, together with a weak evidence
of a barely unbound $s$ or $p$--resonance in $^{10}$Li. A bit later,
Abramovich et al. \cite{abr95} and Zinser et al. \cite{zin95} claimed
that the ground state of $^{10}$Li corresponds to a very low lying
$l$=0 state. In the first case the energies of the doublet 2$^-$/1$^-$
were given to be 27 keV and 88 keV, respectively, while in
\cite{zin95} the energy of the low-lying $s$--state was established
to be below 50 keV.

Recent experiments have been performed in order to clarify this
puzzle.  In \cite{boh97} the existence of a $p$--resonance, probably
the 2$^+$ state, at 0.54$\pm$0.06 MeV has been confirmed, and the
existence of a lower $p$--resonance, the 1$^+$ state, has been found
at 0.24$\pm$0.06 MeV.  Although the existence of a low--lying
$s$--state is not confirmed in this work, they emphasize that part of
the observed strength at the threshold might represent $l$=0 strength.

More recently, Thoennessen et al. \cite{tho99} have repeated the
experiment of Kryger et al. \cite{kry93} with improved energy
resolution in order to establish more stringent limits of the
$s$--wave parameters. They concluded that a low--lying $s$--state was
observed with a scattering length of at least 20 fm corresponding to a
peak energy of less than 50 keV. They also found indications of a
$p$--state around 540 keV, and they did not rule out another $p$-state
at $\sim$240~keV.

In \cite{cag99} the experimental spectrum was fitted with a $p$--state
at $500 \pm 60$~keV and width $400 \pm 60$~keV. Substitution of this
resonance by combination of one $s$ and one $p$ or combination of two
$p$'s provided fits which were less significant due to the increase of
fit parameters. Inclusion of the extra $s$--wave or $p$--wave moved
the 500 keV state up by 15 and 25 keV, respectively. No other states
were identified, either because they were not populated, or because
the experimental resolution was insufficient.  Although there is no
evidence for a state at $\sim$250~keV there is a small enhancement,
also seen in \cite{you94}.

Finally the last experiment by Chartier et al. \cite{cha01} produced
$^{10}$Li by proton knockout of $^{11}$Be. The measured resulting
$^{9}$Li nucleus in the ground state is predominantly produced from
$s$-wave neutron emission of $^{10}$Li which is concluded to resemble
the neutron configuration in the $^{11}$Be ground state. Thus the
ground state of the n-$^{9}$Li system is an $s$-wave below 50~keV.

As a consequence of all this it is well established experimentally
that the ground state of $^{10}$Li corresponds to a very low--lying
$s$--state.  Furthermore a $p$--resonance at an energy of around 0.5
MeV is present in the $^{10}$Li spectrum. The other quantum numbers
specifying the couplings of these states remain so far unknown. Also
the presence of a second $p$--resonance at around 0.25 MeV is not well
established and another $s$-state, which necessarily also must be
present in the low energy spectrum, is entirely unknown.

The observed resonances or virtual states in the neutron-$^{9}$Li
spectrum are related to the second $s_{1/2}$ and the first
$p_{1/2}$-states. The first $s_{1/2}$ and the lowest $p_{3/2}$-states
are occupied in the $^{9}$Li-nucleus. The lowest $s_{1/2}$-state is
strongly bound and the energy of the $p_{3/2}$-state can be estimated
as the one-neutron separation energy of $^{9}$Li, i.e. 4.1~MeV
\cite{sel88,aud95}. Strictly speaking this energy applies to $^{9}$Li
whereas both the $p_{3/2}$ and $p_{1/2}$-states discussed in the
present context are related to the $^{10}$Li-system.

From the theoretical side, initial calculations \cite{tos90}
considered the ground state of $^{10}$Li to be the 1$^+$ state at an
energy of 810 keV, as given in \cite{wil75}. However in \cite{tos90}
is already mentioned the possibility that the ground state of
$^{10}$Li is 2$^-$, as discussed in \cite{bar77}. The first three-body
computation of the $^{11}$Li properties \cite{joh90} predicted that,
to be consistent with the available data, $^{10}$Li should have an
$s$-state below 300 keV in conflict with the 800 keV at that time
believed to be the ground state resonance energy.

In \cite{tho94} it was confirmed that core momentum distributions
after fragmentation of $^{11}$Li were consistent with a low lying
$s$--state in $^{10}$Li, and it was estimated that $^{11}$Li has
nearly 50\% of $s$--wave motion between one neutron and the core. This
fact of a negative parity state was contradicted by subsequent
microscopic calculations \cite{wur96,des97} that predicted a 1$^+$
assignment for the ground state of $^{10}$Li. On the other hand, in
\cite{nun96,vin96} an $s$--state was predicted to be the ground state
of $^{10}$Li.

\subsection{The three-body system }

In \cite{nie01,fed94} a method to solve Faddeev equations in
coordinate space is developed in detail. This method is especially
suitable to describe the large distance behavior of weakly bound
three--body systems. In \cite{gar96,gar97} we have shown that
combination of this method to describe $^{11}$Li and the simple sudden
approximation as fragmentation reaction model are sufficient to
reproduce the shape of the experimental neutron and core momentum
distributions after fast fragmentation of $^{11}$Li on a light
target. We have shown that the agreement with the experiment is
obtained only when the final interaction between fragments is
included, and also when a low--lying $s$--state is present in
$^{10}$Li. Inclusion of only $p$--states in the $^{10}$Li spectrum
necessarily leads to too broad neutron momentum distributions.  

In \cite{gar98} we have investigated the angular correlations after
breakup of $^6$He and $^{11}$Li, and we have shown that the presence
of a low lying $s$--state gives rise to a very different angular
distribution compared to nuclei as $^6$He where the $p$--resonances
dominate in the neutron--core system. The experimental data for the
angular distribution arising from $^{11}$Li was later published
\cite{sim99} and found to be in agreement with the distribution
predicted in \cite{gar98}. This is imperative evidence of strong
$s$--wave configurations in both $^{10}$Li and $^{11}$Li.

More detailed calculations of momentum distributions after
fragmentation of $^{11}$Li on light targets are reported in
\cite{gar99}. The agreement with the experiments is found to be
rather good when around 40\% of $p$--wave is present in the relative
neutron--$^9$Li motion. Nevertheless the precise energies and quantum
numbers of the $^{10}$Li states are not well established, since the
momentum distributions are mainly sensitive to the average energies of
the doublets 1$^-$/2$^-$ and $1^+$/2$^+$ \cite{gar97}. For $^{10}$Li
these states arise from couplings of the $^{9}$Li-spin of 3/2 to the
available $s_{1/2}$ and $p_{1/2}$ neutron states. In $^{11}$Li the
Pauli principle prevents the neutrons from both choosing the lowest
state.  Parity then requires an equal distribution within each
doublet. Thus, most observables are not sensitive to the precise
positions of these doublets, only to their average values. This is the
reason why so many theoretical computations agree fairly well with
experiments even with the obviously wrong assumptions of spin zero for
both $^{9}$Li and $^{11}$Li.  The invariant mass spectrum is an
exception which could be sensitive to the precise position of the
resonances and virtual states.

Very recently the fragmentation model used in \cite{gar99} has been
implemented to take into account the interaction between each of the
halo constituents and the target, as well as to include Coulomb
interaction, making the method valid not only for light targets, but
also for intermediate and heavy targets. This is the
participant--spectator method described in details in
\cite{gar01}. The method works well for $^6$He, where essentially no
uncertainties in the neutron--core interaction are present.

This success for the case of $^6$He leads us to turn the problem
around and investigate in details which structure of $^{10}$Li is
consistent with the firmly established properties of both $^{9}$Li,
$^{10}$Li and $^{11}$Li.  First the neutron-$^{9}$Li potential must be
consistent with the structure of $^{9}$Li and the energies and quantum
numbers of the established two-body resonances and virtual states must
be reproduced.  Then also the known $^{11}$Li three-body structure
must be reproduced. The unavoidable facts are one and only one
particle bound state with angular momentum and parity $\frac{3}{2}^-$,
with binding of 295$\pm$35 keV \cite{you93}, with large interaction
cross sections on all targets and with narrow fragment momentum
distributions after breakup.  The complicated model with finite core
spin is then inescapable.

\subsection{Model dependence }

The measured differential or total $^{11}$Li reaction cross sections
are interpreted as evidence for an exceptionally large spatial
extension of the neutron matter and a correlated wave function mixing
roughly equally relative neutron-core $s$ and $p$-states. The
quantitative details must necessarily be extracted by use of model
computations.  Apart from the binding energy of about 0.30~MeV the
crucial parameter is the $p$-wave content determined to be about 40\%
in a rather successful model \cite{gar99,gar01}, which is able to
describe essentially all measured breakup cross sections.  The breakup
cross sections are first of all sensitive to this $s$-wave content and
we shall therefore aim at reproducing these 60\% in the $^{11}$Li
structure.

The model dependence of the extracted results has to be minimized as
far as possible.  This important issue can be illustrated by two
different examples. The first is the interpretation of the neutron
momentum distribution observed in breakup reactions. Narrower
distributions result from inclusion of both final state interactions
and from removal of those parts of the three-body wave function where
only one or two of the three halo particles appear after the reaction,
i.e. true three-body breakup \cite{gar99,gar01}.  A larger fraction of
neutron-core $s$-wave would compensate and delusively reproduce the
measurement.

The second example is the interaction cross section often attempted to
be expressed in terms of a corresponding value of the $^{11}$Li root
mean square radius.  This interpretation is also model dependent as
evidenced by the different results, i.e. radii of 3.1$\pm$0.3 fm
\cite{tan92}, 3.53$\pm$0.10 fm \cite{alk96} and 3.62$\pm$0.19 fm
\cite{ege01} from Glauber models with parametrized densities,
three-body densities and from elastic proton scattering with gaussian
densities, respectively.

The first of these, off hand least sophisticated, analyses
\cite{tan92} used parametrized static densities of various kinds. The
second analysis used a reaction model accounting for the few-body
correlations obviously present in the initial wave function
\cite{alk96}. The essential input is then the three-body ground state
wave function and the resulting radius relies heavily on the
properties of this choice. The result of about 3.53~fm quoted above
arises from two-body interactions where the virtual states and
resonances cannot all agree with the established experimental
information.  There average positions are unrealistically low and
there is no spin splitting in the calculation.  The $p$-wave content
is almost 60\% and the presently accepted use of a three-body
interaction for fine tuning is not applied.  The third analysis is
from elastic proton scattering resulting in an even larger radius
\cite{ege01}. Again this is one numerical value and corresponding
results for $^{6}$He are in conflict with other information.

Avoiding the model dependence entirely is not possible. Minimizing the
uncertainties at least require that all effects influencing the
observables in question must be accounted for in the model.  The
safest is to use one consistent model with all anticipated physical
effects included and with one set of parameters reproducing all
measured data from the two-body input data to the differential breakup
cross sections. We rely on the three-body structure model \cite{nie01}
and the reaction models developed for systems precisely like $^{11}$Li
\cite{gar99,gar01}.

The paper is organized as follows: In section \ref{sec2} we briefly
sketch the method used to construct the three--body wave function and
the model to compute the breakup cross sections. We continue in
section \ref{sec3} with the details of the interactions involved in
the particular case of $^{11}$Li. In section \ref{sec4} we show
results for the $^{11}$Li wave function and for different relevant
observables after fragmentation reactions. Finally section \ref{sec5}
contains the summary and the conclusions.

\section{Method}
\label{sec2}

The wave function of the three--body halo system is obtained by
solving the Faddeev equations in coordinate space
\cite{nie01,fed94}. This is done by writing each of the three Faddeev
equations in terms of each of the three sets of hyperspherical
coordinates $(\rho,\Omega_i)$, $\Omega_i=(\alpha_i, \Omega_{x_i},
\Omega_{y_i})$, where the index $i$ is related to a given Jacobi
system. The three--body wave function $\Psi^{(JM)}$ is then a sum of
the three Faddeev components, each of them expanded for a given $\rho$
in terms of a complete sets of generalized angular functions
$\Phi_n^{(i)}(\rho,\Omega_i)$

\begin{equation}
\Psi^{(JM)}=\frac{1}{\rho^{5/2}} \sum_n f_n(\rho)\sum_{i=1}^3 
                                       \Phi_n^{(i)}(\rho,\Omega_i) \; .
\end{equation}

The angular wave functions are solutions of the angular part of the
Faddeev equations

\begin{eqnarray}
 {\hbar^2 \over 2m}\frac{1}{\rho^2}\hat\Lambda^2 \Phi^{(i)}_{n}
 +V_{jk} (\Phi^{(i)}_{n}+\Phi^{(j)}_{n} + \Phi^{(k)}_{n}) 
\equiv {\hbar^2 \over
2m}\frac{1}{\rho^2} \lambda_n(\rho) \Phi^{(i)}_{n}  \; ,
\label{eq1}
\end{eqnarray}
where $\{i,j,k\}$ is a cyclic permutation of $\{1,2,3\}$, $m$ is a
normalization mass, $V_{jk}$ is the two-body interaction between
particles $j$ and $k$ and $\hat\Lambda^2$ is the $\rho$-independent
part of the kinetic energy operator. The expression for
$\hat{\Lambda}^2$ is given in \cite{nie01,fed94}.

The radial functions $f_n(\rho)$ are the solutions of 
the coupled set of ``radial'' differential equations
\begin{eqnarray}
   \left(-\frac{\rm d ^2}{\rm d \rho^2}
   -{2m(E-V_3(\rho)) \over\hbar^2} + 
   \frac{ \lambda_n(\rho) }{\rho^2} + \frac{15}{4\rho^2}
 - Q_{n n} \right)   f_n(\rho)  \nonumber \\ 
  = \sum_{n' \neq n}   \left(
   2P_{n n'}{\rm d \over\rm d \rho}
   +Q_{n n'}
   \right)f_{n'}(\rho)  \; ,
\label{eq2}
\end{eqnarray}
where $V_3$ is an anticipated three-body potential and the eigenvalues
of the angular part, $\lambda_n(\rho)$, enter as effective potentials.
The functions $P$ and $Q$ are defined as the angular integrals
\begin{eqnarray}
   P_{n n'}(\rho)\equiv \sum_{i,j=1}^{3}
   \int d\Omega \Phi_n^{(i)\ast}(\rho,\Omega)
   {\partial\over\partial\rho}\Phi_{n'}^{(j)}(\rho,\Omega)  \; , \\
   Q_{n n'}(\rho)\equiv \sum_{i,j=1}^{3}
   \int d\Omega \Phi_n^{(i)\ast}(\rho,\Omega)
   {\partial^2\over\partial\rho^2}\Phi_{n'}^{(j)}(\rho,\Omega)  \; .
\end{eqnarray}

The observables obtained after fragmentation of $^{11}$Li are computed
by using the participant--spectator method \cite{gar01}. The breakup cross 
sections are obtained by adding the incoherent contributions from processes
where one, two, or three halo particles simultaneously interact with
the target. The interaction with the target of one of them (participant) 
is described by a phenomenological optical potential and the
black disk model is used for the other two particles (spectators). 

Let us assume that the halo constituent $i$ is the participant and the
final state can be described as two independent two--body systems, one
made by particle $i$ and the target $0$, and the second made by the
remaining halo constituents $j$ and $k$. This final state description
is appropriate for large momentum transfer processes in which the
constituent $i$ is violently removed from the projectile. Under these
assumptions, and working in the frame of the halo projectile, the
differential cross sections can be written as \cite{gar99,gar01}

\begin{equation}
 \frac{d^9\sigma _{el}^{(i)}(\bd{P}^{\prime },
\bd{p}_{jk}^{\prime },\bd{q})}
{ d\bd{P}^{\prime } d\bd{p}_{jk}^{\prime } d\bd{q} }
  =    \frac{d^3\sigma _{el}^{(0i)}(\bd{p}_{0i}
  \rightarrow  \bd{p}_{0i}^{\prime})}
 {d\bd{q}} \;
\frac{ P_{dis}(\bd{q})}{2 J+1} \sum_{M s_{jk}^{\prime}\Sigma_{jk}^{\prime}
\Sigma_i^{\prime} }
 |M_{s_{jk} \Sigma_{jk}^{\prime} \Sigma_i^{\prime}}^{(JM)}|^{2} \; ,
\label{elas}
\end{equation}

\begin{equation}
 \frac{d^6\sigma _{abs}^{(i)}(\bd{P}^{\prime },\bd{p}_{jk}^{\prime })}
{ d\bd{P}^{\prime } d\bd{p}_{jk}^{\prime } }
=   \sigma _{abs}^{(0i)}(p_{0i}) \;
\frac{1}{2 J+1} \sum_{M s_{jk}^{\prime}\Sigma_{jk}^{\prime}\Sigma_i^{\prime} }
 |M_{s_{jk} \Sigma_{jk}^{\prime} \Sigma_i^{\prime}}^{(JM)}|^{2} \; .
\label{abs}
\end{equation}
where eq.(\ref{elas}) and (\ref{abs}) refer to the cases in which $i$
is elastically scattered and absorbed by the target, respectively.
The cross sections $d^3\sigma _{el}^{(0i)} / d\bd{q}$ and $\sigma
_{abs}^{(0i)}$ are the differential elastic and absorption cross
sections for the participant--target scattering. The momenta
$\bd{P}^\prime$, $\bd{p}_{jk}^\prime$, and $\bd{p}_{0i}^\prime$ are
the relative momenta in the final state between the center of mass of
the two final two--body systems, between the halo constituents $j$ and
$k$, and between $i$ and the target, respectively. The momentum
$\bd{q}$ is the momentum transfer in the process.  The quantum numbers
$(J,M)$ are the total spin and its third component of the halo
nucleus, $(s^\prime_{jk}, \Sigma^\prime_{jk})$ are the spin and third
component of the two--body final state made by particles $j$ and $k$,
and $\Sigma_i^\prime$ is the final state third component of the spin
of particle $i$.  The target is assumed to have spin zero. 

The expressions are then correct when the participant $i$ has spin 0
or 1/2, but we shall employ them even for the spin of 3/2
corresponding to the $^{9}$Li-core.  We expect this is a very accurate
approximation. The function $P_{dis}(\bd{q})$ in eq.(\ref{elas}) is
included to remove the non--zero component of the three--body bound
state plus target in the final state, that represents elastic
scattering of the halo nucleus as a whole. Finally, $M_{s_{jk}
\Sigma_{jk}^{\prime} \Sigma_i^{\prime}}^{(JM)}$ is the overlap between
the $jk$--wave function in the final state and the wave function of
the three--body projectile.

In \cite{gar01} all the details about the participant--spectator model
are given, as well as the optical potentials used to describe the
interaction between participant $i$ and the target.

\section{Two--body interactions and $^{11}$Li wave function}
\label{sec3}

To obtain the $^{11}$Li wave function by use of the procedure sketched
above one needs to specify the two--body interactions entering in
eq.(\ref{eq1}), i.e. the neutron--neutron and the neutron-$^9$Li
interactions.

The neutron-neutron interaction is specified in \cite{gar97b}.  The
neutron--$^{9}$Li interaction is obviously decisive in order to
establish the $^{10}$Li spectrum. We therefore must be very careful in
the choices of both form and parameters. Results from previous work
cannot be used directly since we want, and perhaps even need, a better
correspondence between core-density and potentials of both core and
valence neutrons.

We begin with the core density and potential of the core neutrons.
The $^9$Li-core consists of two neutrons and two protons in the
$s_{1/2}$--shell and four neutrons and one proton in the
$p_{3/2}$--shell.  The mean square radius of $^9$Li is then given by
$\langle r_{^9Li}^2\rangle = \frac{4}{9} \langle r_{s_{1/2}}^2\rangle
+ \frac{5}{9}\langle r_{p_{3/2}}^2\rangle$, where $\langle
r_{s_{1/2}}^2\rangle$ and $\langle r_{p_{3/2}}^2\rangle$ are the mean
square radii of the $s_{1/2}$ and $p_{3/2}$--orbits, respectively.

If we use gaussian potentials for $s$ and $p$--waves with equal range,
$V^{(\ell)}=S^{(\ell)} \exp(-r^2/b^2)$, we can estimate the range of
the interaction by using that i) the root mean square radius of $^9$Li
is 2.32 fm, ii) the lowest $s$--state must be bound and the second
must correspond to a low lying neutron virtual state in $^{10}$Li, and
iii) the $p$--potential must have a bound state with a neutron
separation energy around 4.1 MeV \cite{sel88,aud95}.  The precise
position of the $p_{3/2}$-state is not important in the present work
and we shall use the value 4.1~MeV obtained for $^{9}$Li even though
the potential applies to the neutron-$^{9}$Li system.  The $s_{1/2}$
and $p_{3/2}$ bound states in this potential then produce a $^9$Li
density distribution which strictly speaking is that of the core
particles in $^{10}$Li. This is lack of selfconsistency in line with
an inert core approximation, but only applied to estimate the range of
the potential. This is only of marginal importance provided the
resonances and virtual states remain unchanged.

In table \ref{tab1} we show these potentials fulfilling conditions
ii) and iii) for three different ranges of the gaussian.  In all the
three cases the $s$--potentials have only a bound state and a virtual
state around 300 keV, and the $p$--potential has only a bound state at
$-4.1$ MeV. As we can see in the table, only a range of around 2.0 fm
is giving rise to a root mean square radius for $^9$Li in agreement
with the experimental value of 2.32 fm.

%%%%%%%%%%%%%%%%%%%%%%%%%%%%%%%%%%%%%%%%%%%%
%\end{multicols}
\begin{table}
\caption{Parameters of the trial potentials
$V^{(\ell=0,1)}=S^{(\ell=0,1)} \exp(-r^2/b^2)$ used for the occupied
$s_{1/2}$ and $p_{3/2}$--levels in $^9$Li. The $s$--potential has a
single bound state and a virtual state around 300 keV. The
$p$--potential has a single bound state at $-4.1$ MeV. The root mean
square radius $\langle r_{^9Li}^2\rangle^{1/2}$ of $^9$Li is obtained
from the mean square radii $\langle r_{s_{1/2}}^2\rangle$ and $\langle
r_{p_{3/2}}^2\rangle$ of the $s$ and $p$-orbits, respectively.}
\vspace*{0.2cm}
\begin{center}
\begin{tabular}{|c|ccc|}
\hline
                                           &   b=1.5 fm &  b=2.0 fm  &  b=2.5 fm \\ \hline
$S^{(\ell=0)}$ (MeV)                       &  $-172.0$  &   $-95$    &   $-60$   \\
$\langle r_{s_{1/2}}^2 \rangle^{1/2}$ (fm) &   1.27     &    1.51    &    1.90   \\
$S^{(\ell=1)}$ (MeV)                       & $-142.2$   &  $-86.2$   &  $-59.7$  \\
$\langle r_{p_{3/2}}^2 \rangle^{1/2}$ (fm) &   2.36     &    2.79    &    3.18   \\
$\langle r_{^9Li}^2 \rangle^{1/2}$ (fm)    &   1.95     &    2.31    &    2.69   \\
\hline
\end{tabular}
\vspace*{0.1cm}
\end{center}
\label {tab1}
\end{table}
%\begin{multicols}{2}
%%%%%%%%%%%%%%%%%%%%%%%%%%%%%%%%%%%%%%%%%%%%

We then turn to the valence neutrons, where we take a neutron--$^9$Li
potential with central, spin--spin, and spin--orbit terms
\begin{equation}
V_{nc}^{(\ell)}(r)=V_c^{(\ell)}(r)
   +V^{(\ell)}_{ss}(r) \langle \bd{s_n} \cdot \bd{s}_c \rangle
   +V^{(\ell)}_{so}(r) \bd{\ell}_{nc} \cdot \bd{s}_{n} \; ,
\label{eq7}
\end{equation}
where $\langle \bd{s_n} \cdot \bd{s}_c \rangle= \langle \ell_{nc},
j_{n}, J |\bd{s_n} \cdot \bd{s}_c| \ell_{nc}, j_{n}, J\rangle$,
$\bd{s}_n$ and $\bd{s}_c$ are the intrinsic spins of the neutron and
the core, $\bd{\ell}_{nc}$ is their relative orbital angular momentum,
$j_n$ is the coupled momentum of $\ell_{nc}$ and $s_n$, and $J$ is the
total angular momentum obtained after coupling of $j_n$ and the spin
of the core $s_c$. Each level of the $^{10}$Li spectrum is specified
by the quantum numbers $\left\{ \ell_{nc}, j_n, J\right\}$. In
particular, for $s_{1/2}$--waves ($\ell_{nc}$=0, $j_n$=1/2) the total
angular momentum $J$ can be either 1 or 2. Since the $^9$Li--core has
negative parity the two possible $s_{1/2}$--virtual states are
$J^\pi=1^-$ and $J^\pi=2^-$--states. In the same way there are two
possible $p_{1/2}$--resonances ($\ell_{nc}$=1, $j_n$=1/2) with
$J^\pi=1^+$ and $J^\pi=2^+$. We will refer equally to the two
$s_{1/2}$--virtual states or the doublet $1^-$/$2^-$ and to the
$p_{1/2}$--resonances or the doublet $1^+$/$2^+$.

The radial shapes of the potentials in eq.(\ref{eq7}) are chosen as
gaussians with strengths adjusted independently for each value of
$\ell_{nc}$. Since the range of the gaussians all are equal ($b=2.0$
fm), only the strengths remain as adjustable parameters depending on
orbital angular momentum.  In table \ref{tab2} we show, as function of
strengths, the position of the virtual states for $\ell_{nc}=0$ and
the resonance energies and widths for $\ell_{nc}=1$. Both virtual
states and resonances are obtained as poles of the $S$-matrix. In
previous works we often approximated by using the scattering length
for $s$-waves and values where the phase shift equals $\pi/2$ for
$p$-resonances \cite{gar99}. These energies are systematically
slightly higher than the ones obtained from the $S$-matrix.

%%%%%%%%%%%%%%%%%%%%%%%%%%%%%%%%%%%%%%%%%%%%%%%
%\end{multicols}
\begin{table}
\caption{Strengths ($S$) for the $s$ and $p$ neutron--$^9$Li gaussian
interactions. The energies of the corresponding virtual $s$-states are
$E_s$, and energies and widths of the $p$--resonances are $E_{p}$ and
$\Gamma$. These parameters are precisely defined by real and imaginary
values of the poles of the $S$-matrix.  }
\vspace*{0.2cm}
\begin{center}
\begin{tabular}{|cc|ccc|}
\hline
\multicolumn{2}{|c|}{$\ell_{nc}=0$}& \multicolumn{3}{c|}{$\ell_{nc}=1$} \\
\hline
 $S$ (MeV) & $E_s$ (keV) &  $S$ (MeV) &  $E_{p}$ (MeV) & $\Gamma$ (MeV)    \\
\hline
 $-99.0$    & $65 $      &  $-68.0$   &     0.23             &   0.09            \\
 $-98.0$    & $110$      &  $-67.5$   &     0.30             &   0.14            \\
 $-97.0$    & $168$      &  $-67.0$   &     0.37             &   0.18            \\
 $-96.0$    & $240$      &  $-66.5$   &     0.44             &   0.24            \\
 $-95.0$    & $326$      &  $-66.0$   &     0.50             &   0.30            \\
 $-94.0$    & $428$      &  $-65.5$   &     0.57             &   0.38            \\
 $-93.0$    & $546$      &  $-65.0$   &     0.63             &   0.44            \\
 $-92.0$    & $682$      &  $-64.5$   &     0.69             &   0.52            \\  
 $-91.0$    & $837$      &  $-64.0$   &     0.76             &   0.59            \\
\hline
\end{tabular}
\end{center}
\label{tab2}
\end{table}
%\begin{multicols}{2}
%%%%%%%%%%%%%%%%%%%%%%%%%%%%%%%%%%%%%%%%%%%%%%%

Summarizing, the neutron--$^9$Li interaction is given by
eq.(\ref{eq7}), where the three strengths of the three gaussian
potentials are parameters. For $s$--wave the spin--orbit part does not
contribute and only two parameters remain, the central and the
spin--spin potential. The first determines the average energy of the
1$^-$ and 2$^-$ levels, and the second one is separating these two
states.  For the $p$--waves we start by including only the central and
spin--orbit terms in eq.(\ref{eq7}). One of these two strengths is
determined such that the neutron $p_{3/2}$--state remains at $-4.1$ MeV
corresponding to a strength $V_c^{(\ell=1)}+0.5 V_{so}^{(\ell=1)} =
-86.2$~MeV, as given in table~\ref{tab1}. The $p_{1/2}$--strength
parameter $V_c^{(\ell=1)}-V_{so}^{(\ell=1)}$ can then be used to vary
the resonance energy.  On top of these ordinary central plus
spin-orbit neutron-core terms, the potential in eq.(\ref{eq7})
contains the spin-spin neutron-core interaction (spin-splitting term)
which separates the 1$^+$ and 2$^+$--states and the 1$^-$ and
2$^-$-states.

A very important characteristic of this potential is that the
$s_{1/2}$--interaction (typical strengths are given in
table~\ref{tab2}) has a low--lying virtual state and a deeply bound
state while the $p_{3/2}$--interaction has a bound state at $-4.1$
MeV. Since the $s_{1/2}$--shell and the neutron $p_{3/2}$--shell are
completely filled by the neutrons in the $^9$Li nucleus these states
are forbidden by the Pauli principle when adding more neutrons as for
$^{10}$Li and $^{11}$Li.

We avoid these orbits by use of the phase equivalent potentials
\cite{bay87,coo95} for the $s_{1/2}$ and
$p_{3/2}$--interactions. These potentials have exactly the same phase
shifts as the initial ones, but the bound states are not present.
They decrease exponentially at large distances and diverge as $1/r^2$
at short distances.  In other words, for the $s_{1/2}$--interaction we
take potentials as in table~\ref{tab2}, and for the
$p_{3/2}$--interaction we use the one specified in the third column of
table~\ref{tab1} for $p$--waves.  For use in the three-body
calculation we subsequently, as in \cite{gar99b}, construct their
respective phase equivalent potentials, which finally are used in
eq.(\ref{eq1}).

We have used $l$-dependent potential already by adjusting the
strengths in eq.(\ref{eq7}) independently for each partial wave. It
certainly would have been possible to use $l$-independent potentials
with complicated radial form factors and still reproduce the same
properties. However, the phase equivalent potentials used to account
for the Pauli principle also introduce $l$-dependent potentials. The
simplicity of single gaussian form factors is then preferred since the
potentials in the three-body computation in any case would be
$l$-dependent.

It is well established that use of two--body interactions reproducing
low-energy scattering properties leads to a three--body system that is
marginally underbound compared to experimental values \cite{car98}.
To fine tune the crucial three--body energy we include a three--body
interaction, $V_3(\rho)$, in eq.(\ref{eq2}).  This is a
phenomenological way of accounting for those polarizations of the
particles that are beyond that described by the two--body interactions
\cite{fed96}.  This is off-shell behavior possibly revealed in the
three-body system.  Thus, this interaction must be of short range,
since it only contributes when all three particles interact
simultaneously.

We also use a gaussian shape for this three-body potential.  Since the
hyperradius $\rho$ corresponding to $^9$Li and two neutrons touching
each other is approximately equal to 3.0 fm we use this value as the
range for this three---body force.  Nevertheless the precise value of
the range is not essential and ranges between 2.5 and 4.0~fm all lead
to very similar results.  Another form factor of longer range for
example exponential or falling off with a power law dependence of the
hyperradius would increase the total radius. We maintain the gaussian
shape and leave the strength of the three--body force as a parameter
adjusted to recover the experimental $^{11}$Li binding energy of
$295\pm35$ keV \cite{you93}. In the calculations we shall use 0.30~MeV
when adjusting to the measured binding energy.

\section{Properties of $^{11}$Li}
\label{sec4}

Experimental data and previous theoretical investigations established
that $^{10}$Li has a low--lying virtual $s$--state close to $50$~keV
and a $p_{1/2}$--resonance at around 0.50 MeV.  Furthermore, a
neutron-core $p$--wave content in the $^{11}$Li wave function of
around 40\% is necessary to describe the momentum distributions
obtained after fragmentation of $^{11}$Li.  The unknowns in the
present context are then essentially the energies of the second
$s_{1/2}$--virtual state and the second $p_{1/2}$--resonance. We
adjust the three-body potential to reproduce the crucial binding
energy of about 0.30~MeV.  In this section we investigate which sets
of these $^{10}$Li energies are consistent with the established
properties of $^{11}$Li.

\subsection{Effective radial potentials}

Let us begin by computing the $^{11}$Li wave function without
including the spin--spin term in eq.(\ref{eq7}). This means that the
1$^-$/2$^-$ and the 1$^+$/2$^+$ doublets are degenerate. The energy of
these two doublets corresponds then to the average energy of the
doublet after a subsequent inclusion of the spin-spin potential. In
\cite{gar97} we have shown that the main properties of $^{11}$Li as
well as the behavior of the momentum distributions are determined
basically by these average energies. The most sensitive observable to
the precise positions of the two $s_{1/2}$--virtual states and the two
$p_{1/2}$--resonances is the invariant mass spectrum.

Without inclusion of the spin--spin term the $s_{1/2}$--interaction
has only one free parameter, the strength of the gaussian, used to
place the low--lying virtual $s$--state at the desired value.  Since
the $p_{3/2}$ strength of $-86.2$ MeV is fixed, the $p$--potential
with both central and spin-orbit terms has also only one free
parameter which controls the average energy of the
$p_{1/2}$--resonances.

\begin{figure}
\centerline{\psfig{figure=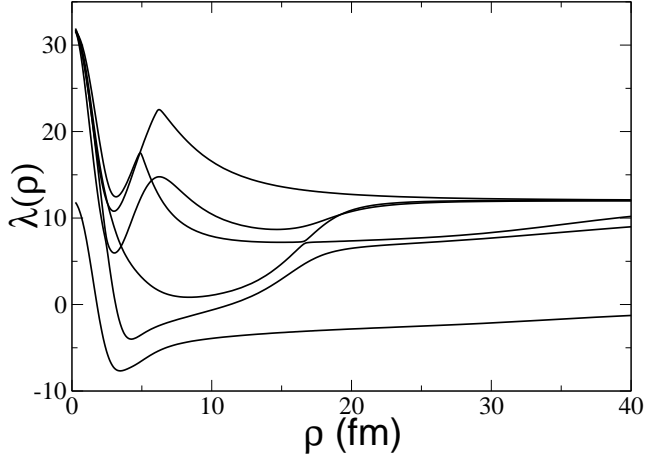,width=8.5cm,%
bbllx=3cm,bblly=1.cm,bburx=19.4cm,bbury=25cm,angle=270}}
\vspace*{0.2cm}
\caption[]{The $\lambda$--spectrum as function of $\rho$ for $^{11}$Li
corresponding to average energies in $^{10}$Li for the virtual
$s$--states and $p_{1/2}$--resonances of around 0.4 MeV. The $s_{1/2}$
and the $p_{3/2}$-potentials are constructed as phase equivalent to
deep potentials but now without any bound states.}
\label{fig1}
\end{figure}

The first step is to compute the angular eigenvalue spectrum by
solving the Faddeev equations in eq.(\ref{eq1}).  We show a typical
example in fig.~\ref{fig1} for $^{11}$Li where all $s$ and $p$--wave
components are included and in addition also all $d$--waves with total
$L$=0 or 1.  The average energies of the 1$^-$/2$^-$ and the
1$^+$/2$^+$ doublets are placed at around 0.4 MeV corresponding to
strengths of $-94.0$ MeV and $-67.0$ MeV for the $s_{1/2}$ and the
$p_{1/2}$--potentials, respectively, see table~\ref{tab2}.  These
eigenvalues define the effective radial potentials, which through
eq.(\ref{eq2}) lead to the radial wave function and the three-body
energy.  Although six eigenvalues are shown in fig.~\ref{fig1}, only
the three lowest are needed to get an accurate $^{11}$Li wave function
and the two lowest already account for more than 98\% of the total
norm. This is intuitively clear from the attractive pockets around
$\rho \approx 4$ fm only appearing for the two lowest eigenvalues in
fig.~\ref{fig1}. We also see that the lowest $\lambda$ is flat, with an
attraction extending to relatively large distances. This is a combined
effect of the large neutron-neutron and neutron-$^{9}$Li $s$-wave
scattering lengths \cite{jen97}.

The $\lambda$--spectrum at the origin and at infinity usually
corresponds to the hyperspherical spectrum $K(K+4)$, where $K$ is the
hypermomentum \cite{nie01,fed94}. A calculation for $^{11}$Li as
described above, but with two--body potentials without bound states
leads to a spectrum with one $\lambda$ starting from 0 ($K=0$) and five
$\lambda$'s from 12 ($K=2$). However, the existence of Pauli forbidden
$s_{1/2}$ and $p_{3/2}$--states and the use of phase equivalent
potentials to exclude them from the three-body calculation is
modifying the $\lambda$--spectrum at short distances \cite{gar99b}. In
fact, as seen in fig.~\ref{fig1}, the $\lambda$ starting at 0, and four
of the $\lambda$'s starting at 12 have disappeared, but the original
spectrum (one $\lambda$ at 0 and five at 12) is preserved at large
distances.

\begin{figure}
\centerline{\psfig{figure=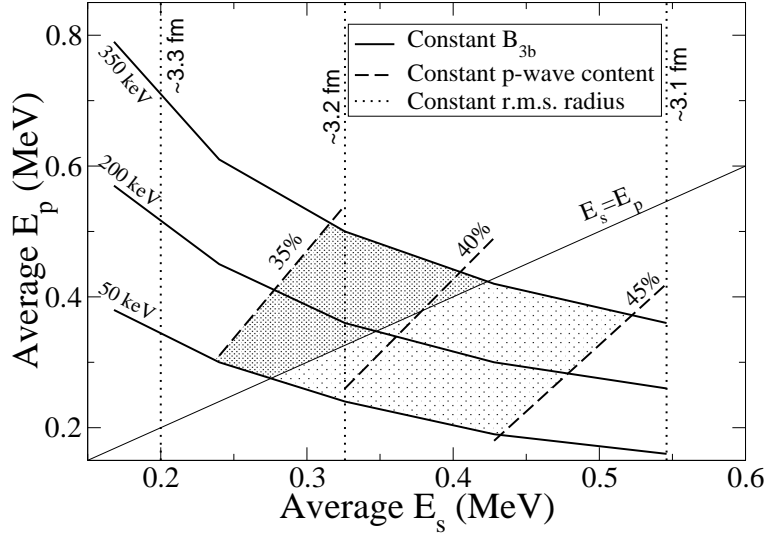,width=10.0cm,%
bbllx=3cm,bblly=1.cm,bburx=19.4cm,bbury=25cm,angle=270}}
\vspace*{0.2cm}
\caption[]{Properties of the $^{11}$Li wave function as function of
the statistically weighted average energies of the two virtual
$s_{1/2}$--states ($E_s$) and the two $p_{1/2}$--resonances
($E_p$). The total three-body binding energy is 0.30 MeV consistent
with the measurement. Along the solid lines we have constant binding
energy contributions $B_{3b} \equiv - \langle V_3(\rho) \rangle$ of
the expectation value of the three--body force for the corresponding
solutions. Along the dashed lines the $p$--wave content in the
$^{11}$Li wave function is constant and along the dotted lines the
root mean square radius is constant. The thin solid line indicates
when the average $s_{1/2}$ and $p_{1/2}$--energies are equal.  The
shaded regions between the lines labeled 35\%, 45\%, 50~keV and
350~keV, indicate the possible values for the average $s_{1/2}$ and
$p_{1/2}$--energies. }
\label{fig2}
\end{figure}

\subsection{Variation with spin averaged parameters}

The statistically weighted average values of the two
$s_{1/2}$-energies and correspondingly of the two $p_{1/2}$-energies
are the crucial quantities. The individual positions are only
important for very specific observables. In the following we shall
refer to these combinations simply as average values.  The two free
parameters are now used to vary these average energies of the
(unbound) $s_{1/2}$ and $p_{1/2}$-states in $^{10}$Li. The resulting
properties of the $^{11}$Li wave function are shown in
fig.~\ref{fig2}. Selecting a set of these two parameters does not
automatically produce a bound $^{11}$Li system. In any case the
expectation value $-B_{3b} \equiv \langle V_3(\rho) \rangle$ of the
three-body potential, where the wave function used in the expectation
value is that of the three-body solution, must vary with the
parameters. The total energy is decisive for many properties and we
therefore adjust the three--body force, $V_3(\rho)$, in eq.(\ref{eq1})
to give a total three-body binding energy of 0.30 MeV for
$^{11}$Li. Then the strength of the gaussian $V_3(\rho)$ potential is
a function of the $^{10}$Li parameters.  We show the curves of
constant three--body binding energy corresponding to $B_{3b}$ 50 keV,
200 keV and 350 keV.  Almost independent of $E_p$ we obtain the curves
of constant radii, i.e. 3.1 fm, 3.2 fm and 3.3 fm. The lines of
constant $p$-wave content, 35\%, 40\%, and 45\%, must correspond to
simultaneous increase of the weighted average values of both $s$ and
$p$-energies.  From fig.~\ref{fig2} we can already extract relevant
information:

\begin{itemize}
\item[-] Values of the average $p_{1/2}$--energy, giving rise to a
$^{11}$Li structure with a $p$--wave content around 40\% compatible
with the known fragmentation data, are always below 0.5~MeV.  Energies
above this value require a large energy contribution from the
three--body force, and also a rather high value of the average
$s_{1/2}$--energy.  Therefore the lowest $p_{1/2}$--state in $^{10}$Li
can not be above 0.5--0.6 MeV.

\item[-] In the same way, reasonable values of the average
$p_{1/2}$--energy are always above 0.2 MeV. Lower values would lead to
a too bound $^{11}$Li wave function, or at least with a very little
effect of the three--body force. The possibility of very high values
of the $s_{1/2}$--energy leads to a too small r.m.s. radius
substantially below 3.1 fm and would in addition require essentially
zero spin-splitting to avoid binding of $^{10}$Li. Therefore 0.2 MeV
is the lowest limit for the average $p_{1/2}$--energy, implying that
the highest $p_{1/2}$--resonance has to be above this value.

\item[-] The values of the average $s_{1/2}$--energy, giving a
$^{11}$Li with reasonable size, binding and $p$--wave content, range
between 250 keV and 550 keV.

\item[-] An average $s_{1/2}$--energy below the average
$p_{1/2}$--energy requires $p$-wave contents between 35\% and
40\%. Higher $p$-wave contents need the $s_{1/2}$--energy above the
$p_{1/2}$--energy.

\item[-] A $p$--wave content around 50\%, as suggested in
\cite{tho94}, needs an average $p_{1/2}$--energy below 0.3--0.4 MeV,
and a significantly higher average $s_{1/2}$--energy close to
600~keV. In any case the size of such $^{11}$Li would be clearly
smaller than 3.1 fm, the lower limit of the radius.

\item[-] A r.m.s. larger than 3.3 fm produces a $^{11}$Li with too
little $p$--wave content, not higher than 25\%.

\item[-] Therefore, the regions of possible values for the average
$s_{1/2}$ and $p_{1/2}$--energies are the shaded regions in
fig.~\ref{fig2}, where we assumed that $^{11}$Li has a $p$--wave
content between 35\% and 45\%, that the limits of the r.m.s. radius of
$^{11}$Li are 3.2$\pm$0.1 fm and finally that the three--body
computations need a finite three--body force.  The dark shade
corresponds to average $s$--energies below the $p_{1/2}$--energies,
while the lighter area is the other way around.  There is no
requirement of an average $s_{1/2}$--energy below that of $p_{1/2}$.  A
sufficiently large spin--splitting in the $s_{1/2}$--interaction could
produce the experimentally required very low--lying $s_{1/2}$--virtual
state and simultaneously maintain the average $s$--energy above the
average $p_{1/2}$-energy.

\item[-] One of the most firmly established properties of the
$^{10}$Li spectrum is the existence of a $p_{1/2}$--resonance at
around 0.5 MeV. According to fig.~\ref{fig2} another
$p_{1/2}$--resonance must be present at an energy below 0.5 MeV in
order to maintain the average energy in the shaded region.  The only
experimental evidence for this low-lying $p$-resonance is the Berlin
group \cite{boh97} giving an energy of 0.24$\pm$0.06 MeV.

\end{itemize}

\begin{figure}
\centerline{\psfig{figure=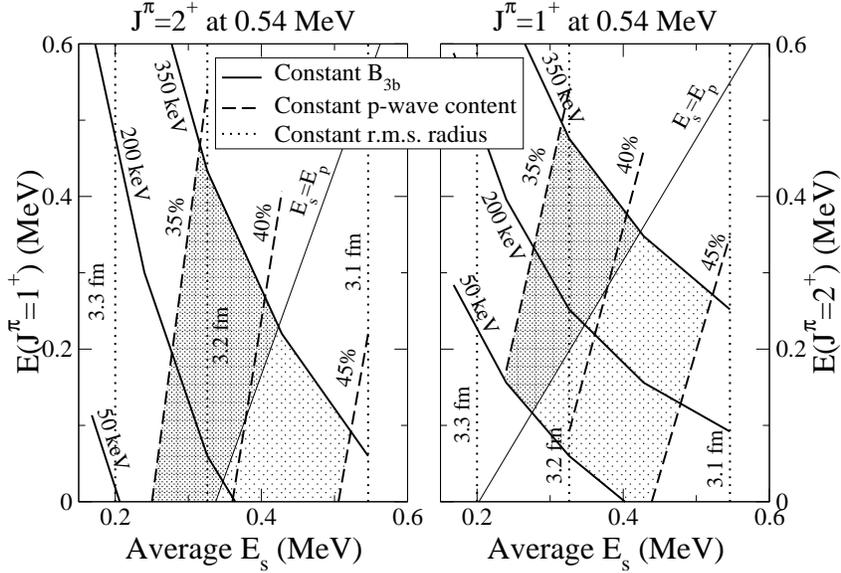,width=10.0cm,%
bbllx=1cm,bblly=1.cm,bburx=17.4cm,bbury=25cm,angle=270}}
\vspace*{1.2cm}
\caption[]{Left: Properties of the $^{11}$Li wave function as function
of the average energy of the two virtual $s$--states ($E_s$) and the
energy of the 1$^+$ $p$--resonance when the 2$^+$ $p$--resonance is
kept at 0.54 MeV. Right: The same as the left part but now as function
of the energy of the 2$^+$ $p$ --resonance with the 1$^+$
$p$--resonance fixed at 0.54 MeV The meaning of the curves is as in
fig.~\protect\ref{fig2}.}
\label{fig3}
\end{figure}

\subsection{Effects of spin splitting}

Let us now in the potential (\ref{eq7}) consider the spin--spin
term which also is assumed to be a gaussian with the same range
($b=2.0$ fm) as the central and spin--orbit terms.  Inclusion of this
term in the $p$--potential separates the two $p_{1/2}$--resonances and
the strength of the gaussian can then be used to split the energy of
these resonances.  We show in fig.~\ref{fig3} the same kind of plot as
in fig.~\ref{fig2}, but now with a fixed energy of 0.54 MeV for one of
the $p_{1/2}$--resonances. Thus we show the properties of $^{11}$Li as
function of the average $s_{1/2}$--energy and the 2$^+$ (left part) or
1$^+$ (right part) $p$-resonances fixed at 0.54 MeV.  Again the
shaded areas correspond to the region consistent with the
established properties of $^{11}$Li.

In both parts of the figure we confirm clearly that a
$p_{1/2}$--resonance at 0.54 MeV necessarily needs a second
$p_{1/2}$--resonance at a lower energy.  If the resonance,
$J^\pi=2^+$, is at 0.54 MeV (left part) then the $J^\pi=1^+$ resonance
can occur at very low values.  In fact, if we consider $^{11}$Li with
40\% $p$--wave content and a binding energy contribution of the
three--body force of 200 keV then the $1^+$--resonance would be very
close to threshold.  A $1^+$-state at 0.24 MeV (as suggested in
\cite{boh97}) would require either a $p$-wave content smaller than
40\% or a large contribution to the energy of the three--body force.
The maximum value for the 1$^+$--resonance is around 0.4 MeV.

\begin{figure}
\centerline{\psfig{figure=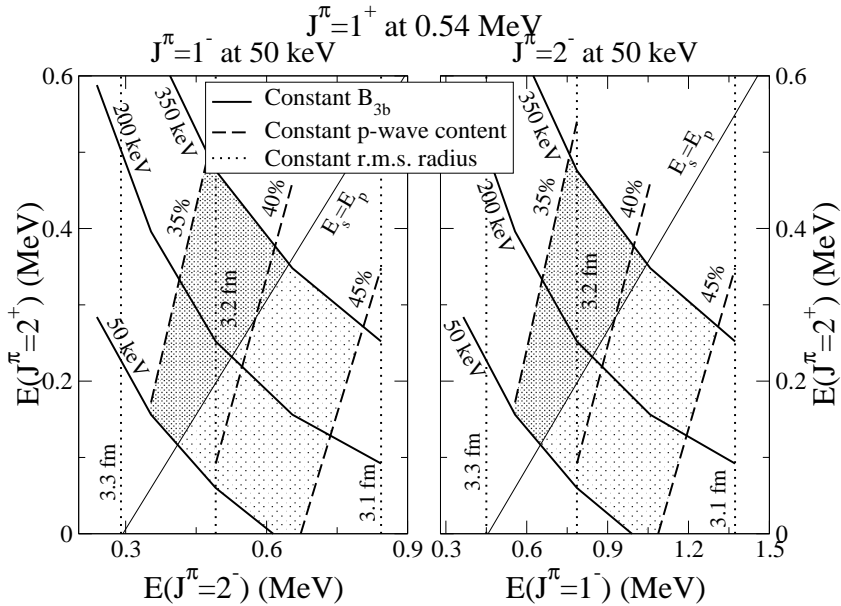,width=10.0cm,%
bbllx=1cm,bblly=1.cm,bburx=17.4cm,bbury=25cm,angle=270}}
\vspace*{1.1cm}
\caption[]{Left: Properties of the $^{11}$Li wave function as function
of the energies of the $2^-$ and the 2$^+$-energies when the
1$^+$--resonance is at 0.54 MeV and the $1^-$ is at 50 keV. Right: The
same as the left part but now as as function of the 1$^-$ energy with
a fixed 2$^-$--state at 50 keV.}
\label{fig4}
\end{figure}

If the $J^\pi=1^+$ resonance is at 0.54 MeV (right part) then the
2$^+$-state can also occur at low values, although in this case
resonances close to threshold have a $p$--wave content close to 45\%
and a smaller contribution to the energy from the three--body force.
Then a $^{11}$Li structure with around 40\% $p$--wave and binding
energy contribution of the three--body force close to 200 keV is
consistent with a second $p_{1/2}$--resonance at 0.24$\pm$0.06
MeV. The maximum values for the 2$^+$ are still around 0.4 MeV.

In figs.~\ref{fig2} and \ref{fig3} we considered average energies for
the $s_{1/2}$--states and we concluded that values between 250 and 550
keV are in agreement with the known properties of $^{11}$Li.
Experimentally the existence of a very low--lying $s$--state close to
50 keV is known.  We then use the spin--splitting term of the
potential for $s$--waves to fix one of the two $s_{1/2}$--virtual
states at 50 keV.  In fig.~\ref{fig4} we fix the 1$^+$--resonance at
0.54 MeV (as in the right part of fig.~\ref{fig3}) and fix one of the
$s_{1/2}$--states at 50 keV (the 1$^-$ in the left part and the 2$^-$
in the right part).  We then plot the properties of $^{11}$Li as
function of the 2$^+$ and 2$^-$ energies in the left part of the
figure and as function of the 2$^+$ and 1$^-$ energies in the right
part of the figure.

\begin{figure}
\centerline{\psfig{figure=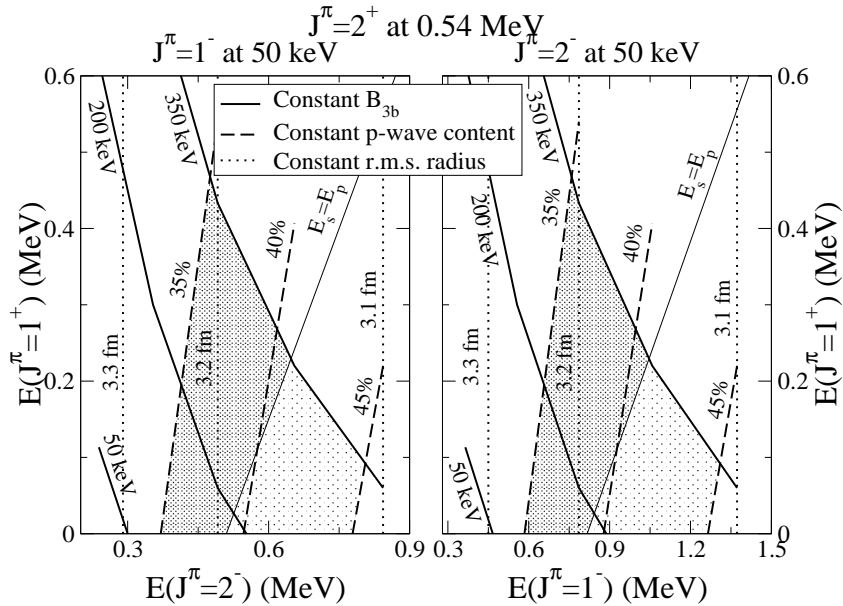,width=10.0cm,%
bbllx=1.cm,bblly=1.cm,bburx=17.4cm,bbury=25cm,angle=270}}
\vspace*{1.1cm}
\caption[]{The same as fig.~\ref{fig4} but fixing the $2^+$ energy at
0.54 MeV and plotting the $^{11}$Li properties as function of the
$1^+$ energy.}
\label{fig5}
\end{figure}

We see that inclusion of the spin--splitting potential for $s$--waves
does not change the conclusions about the $2^+$ energy compared to the
left part of fig.~\ref{fig3}, and the only new information is that if
the 2$^-$ state is at 50 keV, then the 1$^-$--energy moves between 0.6
and 1.2 MeV. On the other hand, if the virtual state at 50 keV is the
1$^-$-state then the second $s$--state is at an energy between 400 and
750 keV.  The same is observed in fig.~\ref{fig5}, constructed in
analogy to fig.~\ref{fig4}, but with the $2^+$--resonance at 0.54
MeV. Then fig.~\ref{fig5} corresponds to the left part of
fig.~\ref{fig3} after including the spin--splitting term in the
$s$--potential.

\subsection{The $^{10}$Li spectrum}

\begin{figure}[t]
\centerline{\psfig{figure=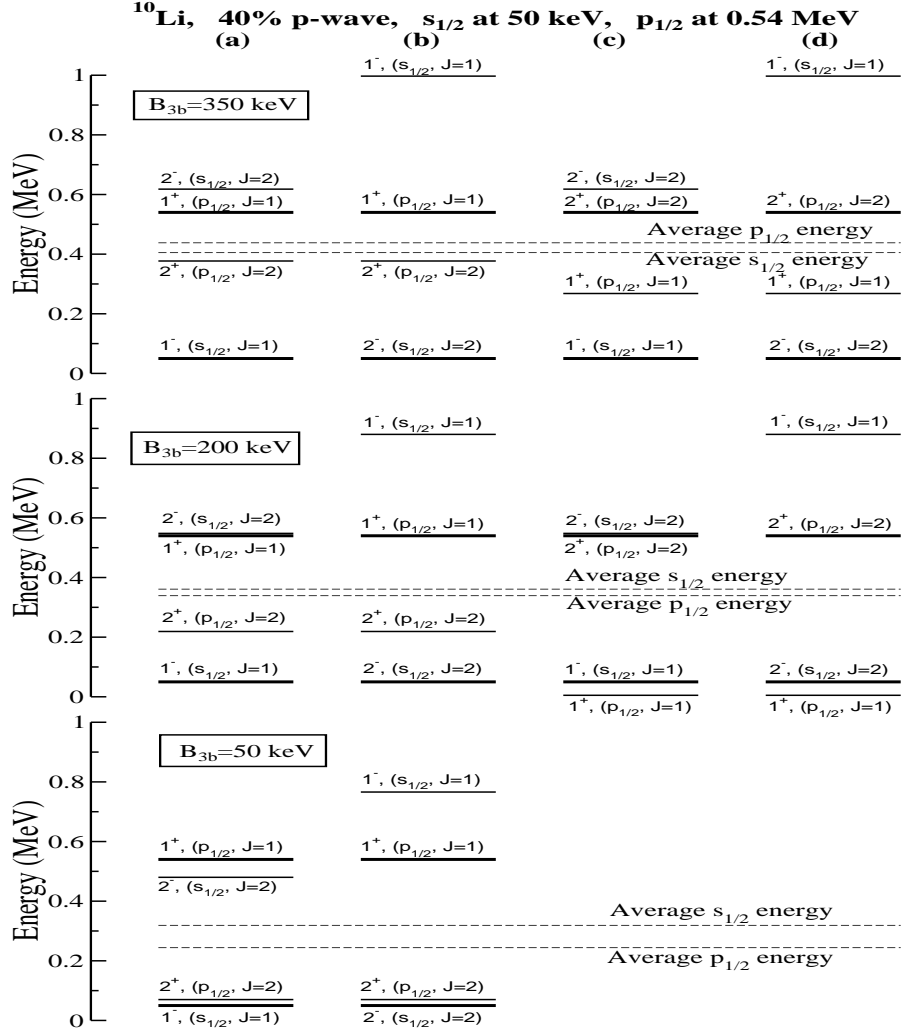,height=16.0cm,width=12.0cm,%
bbllx=6.5cm,bblly=5.3cm,bburx=15.2cm,bbury=23.2cm,angle=0}}
\vspace*{-2.0cm}
\caption[]{ Spectra for $^{10}$Li assuming an
$s_{1/2}$--virtual state at 50 keV, a $p_{1/2}$--resonance at 0.54
MeV, and a $p$--wave content of 40\%. The upper, central, and lower
parts show the cases where $B_{3b}=350$ keV, 200 keV, and 50 keV,
respectively. Columns (a), (b), (c), and (d) correspond to the cases
shown in the left part of fig.~\ref{fig4}, right part of
fig.~\ref{fig4}, left part of fig.~\ref{fig5}, and right part of
fig.~\ref{fig5}, respectively. Levels indicated with a thick line are
the $s$ and $p$--levels for which the energy have been fixed at 50 keV
and 0.54 MeV, respectively.}
\label{fig6}
\end{figure}

The information obtained from figs.~\ref{fig4} and \ref{fig5} can now
be used to predict the spectrum of $^{10}$Li for different properties
of $^{11}$Li while assuming a $p_{1/2}$--resonance at 0.54 MeV and a
low--lying $s_{1/2}$--state at 50 keV.  For instance, if we assume a
$p$--wave content of 40\% and a 350 keV for $B_{3b}$ the energies of
the remaining two $^{10}$Li states are determined where these curves
cross each other.  From the left part of fig.~\ref{fig4} we find an
energy for the 2$^+$ state close to 0.4 MeV, and a bit above 0.6 MeV
for the 2$^-$--state.  The resulting spectrum of $^{10}$Li for this
case is shown in column (a) of the upper part of fig.~\ref{fig6}.
From the right part of fig.~\ref{fig4} we find that the 2$^-$state is
at 50 keV and the 1$^-$-state is then close to 1 MeV.  This spectrum
is shown in column (b) of the upper part of fig.~\ref{fig6}.

In the same way, assuming a $p$--wave content of 40\% and $B_{3b}=350$
keV, we obtain from the left and right part of fig.~\ref{fig5} the
spectra shown in columns (c) and (d) of fig.~\ref{fig6}.  We observe
that these two cases are consistent with the 1$^+$--level at
0.24$\pm$0.06 MeV and the 2$^+$--level at 0.54$\pm$0.06 MeV observed
in \cite{boh97}.  The average energy for the $s_{1/2}$--levels is
close to 0.4 MeV in all the four cases, and it is a bit below the
value around 0.45 MeV of the $p_{1/2}$--levels.

Although a $p$--wave content of 40\% is consistent with previous
calculations, the energy given by the three--body force is not
known. In the central part of fig.~\ref{fig6} we show the same spectra
for $^{10}$Li as in the upper part but assuming $B_{3b}=200$ keV.  The
four spectra correspond to the crossing between the $B_{3b}=200$ keV
line and the line showing a $p$--wave content of 40\% in the left and
right parts of figs.~\ref{fig4} and \ref{fig5}, respectively.  Now the
average energy of the $p$--resonances is slightly lower than the
energy of the $s$--states and both of them are smaller than those from
the upper part. This is because the two--body interactions give more
binding than in the previous case.

\begin{table}
\caption{Strength parameters in MeV of the two-body neutron-$^{9}$Li
potentials of the form $V^{(\ell)}_{i}=S^{(\ell)}_{i} \exp(-r^2/b^2)$,
where $i = c, ss, so$, see eq.(\ref{eq7}). The strengths of the
three-body potential of the form $V_{3b} = S_{3b} \exp(-r^2/b_3^2)$
are given in MeV. The first column gives the sequence of quantum
numbers of the spectrum in fig.~\ref{fig6} with the lowest first.
The ranges are $b=2.0$ fm and $b_{3}=3.0$ fm
in all cases. The 10 different sets correspond to the spectra of
fig.~\ref{fig6}, where we start in the upper left corner taking line
by line.  }
\vspace*{0.3cm}
\begin{center}
\begin{tabular}{|c|cc|ccc|c|}
\hline
 case  & $S^{(\ell=0)}_{c}$ & $S^{(\ell=0)}_{ss}$ & $S^{(\ell=1)}_{c}$ & 
$S^{(\ell=1)}_{so}$ & $S^{(\ell=1)}_{ss}$ & $S_{3b}$  \\
\hline
 $1^-2^+1^+2^-$ & $-94.0$ & 6.85    & $-79.64$ & $-13.12$ & $-0.65$ & $-3.9$   \\

 $2^-2^+1^+1^-$ & $-94.0$ & $-11.4$ & $-79.64$ & $-13.12$ & $-0.65$ & $-3.9$   \\

 $1^-1^+2^+2^-$ & $-94.0$ & 6.85    & $-79.64$ & $-13.12$ &   1.10  & $-3.9$   \\

 $2^-1^+2^+1^-$ & $-94.0$ & $-11.4$ & $-79.64$ & $-13.12$ &   1.10  & $-3.9$   \\
\hline
 $1^-2^+1^+2^-$ & $-94.6$ & 6.10    & $-79.86$ & $-12.68$ & $-1.20$ & $-2.3$   \\

 $2^-2^+1^+1^-$ & $-94.6$ & $-10.1$ & $-79.86$ & $-12.68$ & $-1.20$ & $-2.3$   \\

 $1^+1^-2^+2^-$ & $-94.6$ & 6.10    & $-79.86$ & $-12.68$ &    1.95 & $-2.3$   \\

 $1^+2^-2^+1^-$ & $-94.6$ & $-10.1$ & $-79.86$ & $-12.68$ &    1.95 & $-2.3$   \\
\hline
 $1^-2^+2^-1^+$ & $-95.0$ & 5.55    & $-80.08$ & $-12.25$ & $-1.70$ & $-0.8$   \\

 $2^-2^+1^+1^-$ & $-95.0$ & $-9.3$  & $-80.08$ & $-12.25$ & $-1.70$ & $-0.8$   \\
\hline
\end{tabular}
\vspace*{0.2cm}
\end{center}
\label {tab3}
\end{table}

Columns (a) and (b) of these spectra can also be consistent with a
second $p_{1/2}$--resonance around 0.24 MeV, although now this
resonance corresponds to the 2$^+$--level instead of the
1$^+$--level as given in \cite{boh97}. In the spectra (c) and (d) in
the central part of fig.~\ref{fig6} the lowest $p_{1/2}$--resonance is
very close to threshold (the crossing between $B_{3b}=200$ keV and
$p$--wave content of 40\% in fig.~\ref{fig5} is very close to zero
$1^+$--energy). As previously indicated there is no experimental
evidence of such a low $p_{1/2}$--resonance, and it is established
that the ground state corresponds to an $s_{1/2}$--state.

Finally, in the lower part of fig.~\ref{fig6} we show the $^{10}$Li
spectrum by assuming $B_{3b}=50$ keV. Again, since the three--body
force is providing less binding the average energy of the $s$--states
and the $p$--resonances is smaller than in the previous cases. The
role played by the three--body force is now not very important.  The
crossing between the lines corresponding to constant $B_{3b}$=50~kev
and constant $p$--wave content of 40\% in fig.~\ref{fig4} gives rise to
the two spectra shown in the figure.  In this case also a
$p_{1/2}$--resonance (the 2$^+$--state) appears at an energy below 100
keV.

From fig.~\ref{fig5} the same procedure gives negative energies of
the 1$^+$--level, meaning that this level is bound. This is of course
against the well known fact that $^{10}$Li is not bound, and the
corresponding $^{10}$Li spectrum can certainly not correspond to the
real one. Actually, from the spectra (c) and (d) in the middle part of
fig.~\ref{fig6}, we already see that a value of $B_{3b}$ slightly below
200 keV is already binding the $1^+$--level.

For completeness we give in table~\ref{tab3} the two and three--body
potentials corresponding to all the spectra shown in fig.~\ref{fig6}.

\subsection{Invariant mass spectra in fragmentation reaction}

According to the experimental and theoretical information collected
about $^{10}$Li and $^{11}$Li we can conclude that the spectra shown
in the upper part of fig.~\ref{fig6}, especially in (c) and (d), and
spectra (a) and (b) shown in the middle part, are the main candidates
for the $^{10}$Li--spectrum. From the different observables obtained
experimentally after fragmentation of $^{11}$Li the invariant mass
spectrum is the observable most sensitive to the energies of the
different resonances and virtual states. Investigation of the
invariant mass spectrum could then help to establish which spectrum is
most probable for $^{10}$Li.

\begin{figure}[t]
\centerline{\psfig{figure=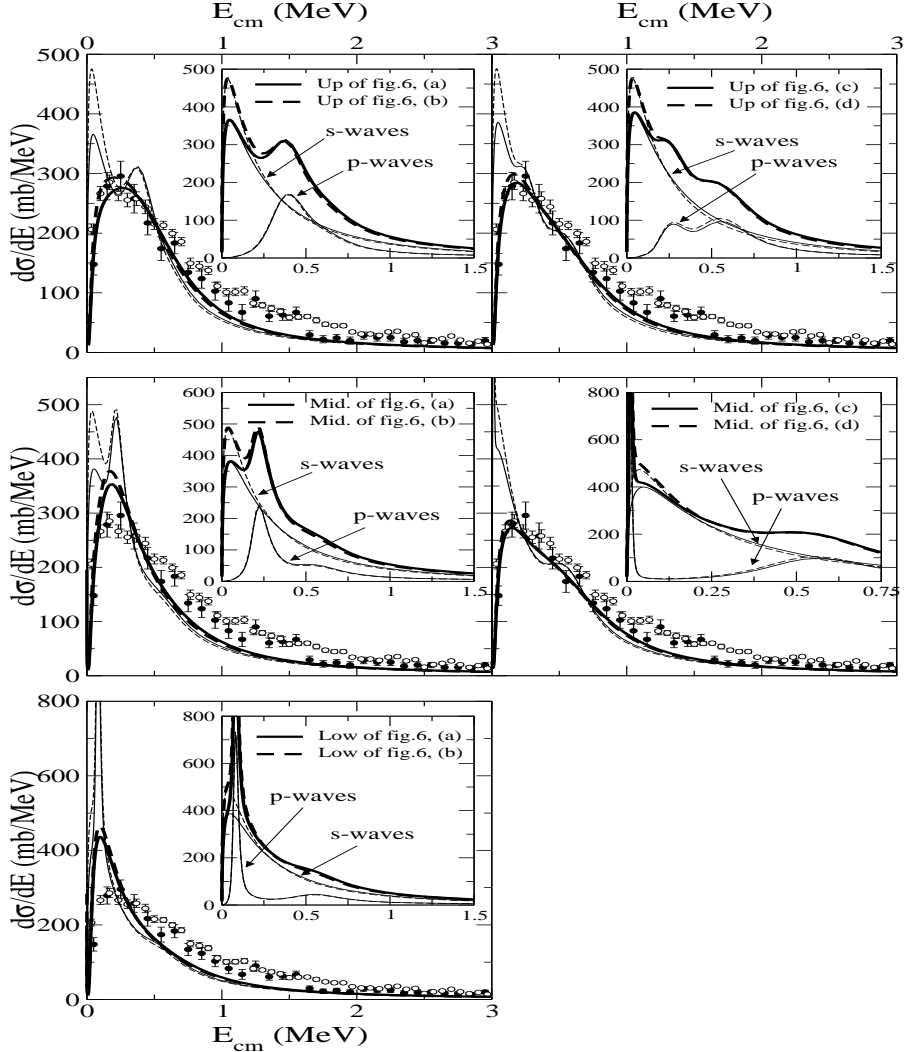,height=16cm,width=12cm,%
bbllx=6.5cm,bblly=5.3cm,bburx=15.2cm,bbury=23.2cm,angle=0}}
\vspace*{-2cm}
\caption[]{Invariant mass distributions obtained with the different
$^{10}$Li spectra shown in fig.~\ref{fig6}. The upper, central, and
lower parts correspond to the upper, central, and lower parts of
fig.~\ref{fig6}. Distributions in the left panels correspond to the
$^{10}$Li spectra (a) solid lines, and (b) dashed lines of
fig.~\ref{fig6}, while the distributions in the right panels correspond
to the $^{10}$Li spectra (c) solid lines, and (d) dashed lines of
fig.~\ref{fig6}. In the main plots we show the invariant mass
distributions for each case and the results obtained after convoluting
with the experimental beam profile. The internal plots show the
invariant mass distributions and the contribution to the total from
the $s$--waves and the $p$--waves. Experimental data are from
\cite{zin97} filled circles, and from \cite{sim98} open circles.}
\label{fig7}
\end{figure}

In fig.~\ref{fig7} we show the invariant mass spectrum for all the
$^{10}$Li spectra shown in fig.~\ref{fig6}. The observables after
fragmentation of $^{11}$Li are computed following the
participant--spectator method described in \cite{gar01}.  The external
parts of each graph in the figure show the computed invariant mass
spectrum (thin lines) and the result after convoluting with the
experimental beam profile (thick lines). This curve is compared with
the experimental data taken from \cite{zin97} and \cite{sim98}. These
sets of experimental data do not agree in the absolute numbers, and we
have scaled those of \cite{sim98} to the same maximum as in
\cite{zin97}.

Note that the only difference between the $^{10}$Li--spectra (a) and
(b) and between (c) and (d) is the ordering of the $1^-$ and the
$2^-$--levels. Therefore the compared solid lines and dashed lines in
fig.~\ref{fig7} differ only in the ordering of the
1$^-$/2$^-$--doublet. For the solid lines the $1^-$--state is at 50
keV, while for the dashed lines the 2$^-$--state is at 50 keV.

As we can see in all the panels in fig.~\ref{fig7} the difference
between solid and dashed lines is very small, especially after
convoluting with the experimental beam profile. The only difference
appears before convoluting and at very small energies, where the cases
with the 2$^-$--state at 50 keV (dashed lines) have a higher
peak. This is due to the larger statistical weight of the $2^-$--level
compared to the 1$^-$--level (5/3 times bigger).  In any case, from
the invariant mass distributions and its comparison with the present
experimental data it is not possible to determine which of the two
states in the doublet $1^-$/$2^-$ corresponds to the low--lying
$s$--state found experimentally.

Let us concentrate now on the $^{10}$Li--spectra (a) and (b). In the
three cases, upper, middle, and lower parts, one of the $s$--states is
placed at 50 keV, and the $1^+$--resonance is fixed at 0.54
MeV. Basically the only difference between the three cases is the
energy of the 2$^+$--level, that goes from almost 0.4 MeV to a small
value below 100 keV. Looking at the corresponding invariant mass
distributions in the left part of fig.~\ref{fig7} we observe that
calculations before convoluting with the experimental beam profile
show two peaks corresponding to the low $s$--state at 50 keV, and the
$2^+$--resonance. These two peaks approach each other when the energy
of these two levels becomes closer, and the one corresponding to the
$2^+$--resonance becomes higher when its energy approaches the
threshold. In fact, in the lower part of fig.~\ref{fig7} both peaks can
not be distinguished. 

The peak corresponding to the $1^+$--state at 0.54 MeV can not be
seen.  Looking at the $p$--wave contribution (inner panels) we note
that only in the lower part this peak can be noticed in the small bump
produced in the total invariant mass distribution, although this bump
disappears after convolution with the beam profile.  The main effect
produced by the different energies of the $2^+$--resonance is that the
lower the energy of the state the narrower the computed invariant mass
spectrum.  This fact is preserved after convoluting the computed
distributions with the experimental beam profile. The main effect of
the convolution is that the separated peaks associated to the
low--lying $s$--state and the $2^+$--resonance merge into one. In any
case the comparison with the experimental distribution shows that the
computed ones are too narrow, except perhaps for values of the
$2^+$--level close or higher than 0.4 MeV.  This would mean that the
two states in the $1^+$/$2^+$--doublet would be almost degenerate.
Therefore a $2^+$--level below the $1^+$--level is not likely.

Let us focus now on the $^{10}$Li--spectra (c) and (d). Again one of
the $s$--levels is placed at 50 keV, but now the $2^+$--level is fixed
at 0.54 MeV, while the energy of the $1^+$--level takes different
values. In the upper part the 1$^+$--resonance has an energy of 0.25
MeV. The computed invariant mass spectrum shows two shoulders produced
by the two $p$--resonances. These two states are clearly seen in the
$p$--wave contribution shown in the inset of the upper part of
fig.~\ref{fig7}.  Nevertheless the structure observed in the computed
invariant mass spectrum disappears after convolution. The agreement
with the experimental data is quite satisfactory, especially with the
one given in \cite{zin97}.  If we reduce the energy of the
$1^+$--level its contribution appears in a higher and narrower peak.

In the middle part of fig.~\ref{fig7} we show the extreme case in which
the $1^+$--level is very close to the threshold (its energy is around
10 keV). As we see in the inset, the $p$--wave contribution has a wide
peak at 0.5 MeV and a sharp and very high peak at a very low energy
corresponding to the 1$^+$--resonance. This is obviously producing a
similar high and narrow peak in the computed invariant mass
spectrum. Nevertheless after convolution this peak is not visible
anymore, and the comparison with the experiment can also be considered
rather good. This is because although convolution is diluting the
presence of the sharp peak at low energies there is still a resonance
at 0.54 MeV, with high statistical weight, that makes the tail of the
total distribution behave accordingly. Therefore a $1^+$--level below
the $2^+$ is the most likely structure for the $^{10}$Li--spectrum.
This result agrees with \cite{boh97}.

From the simultaneous analysis of figs.~\ref{fig6} and \ref{fig7} we
can then summarize the results in the following points: i) We can not
establish which of the two states in the doublet $1^-$/$2^-$ is the
one at a very low energy, ii) a $2^+$ resonance below the $1^+$--level
is not likely, producing too narrow invariant mass spectra, unless the
$1^+$ and the $2^+$--states are close to degeneracy, iii) the most
likely situation corresponds to a $1^+$--level below the $2^+$--state,
although it can not be extracted by direct comparison with the
experimental invariant mass spectra, (iv) experiments with a
better beam energy resolution are desirable, since the predicted
structure in the invariant mass spectra then could be detected and the
results from the different spectra directly distinguished.

\subsection{Momentum distribution after fragmentation}

\begin{figure}[t]
\centerline{\psfig{figure=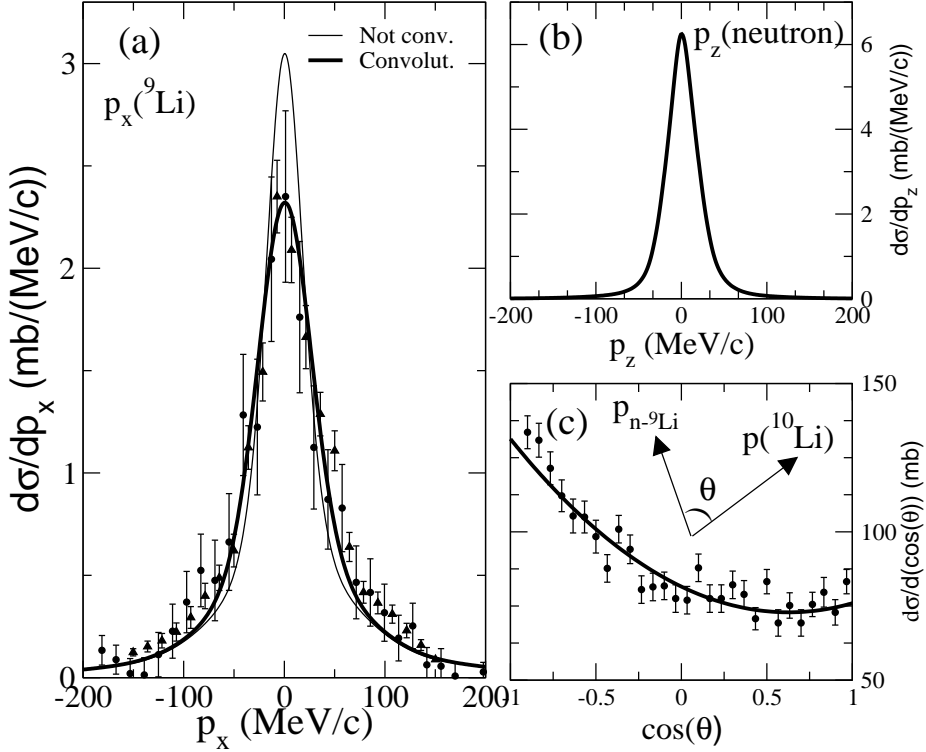,width=12cm,%
bbllx=2.0cm,bblly=1.6cm,bburx=21.3cm,bbury=24.4cm,angle=270}}
\vspace*{0.1cm}
\caption[]{ (a) core momentum distribution, (b) parallel neutron
momentum distribution, and (c) angular distribution after
fragmentation of $^{11}$Li on carbon at 280 MeV/nucleon. In (a) the
thin line is the computed curve and the thick line is the distribution
obtained after convolution with the experimental beam profile. The
experimental data are taken from \cite{hum95,gei} in (a) and from
\cite{sim99} in (c). }
\label{fig8}
\end{figure}

As a final test, we show in fig.~\ref{fig8} some additional
observables after fragmentation of $^{11}$Li on carbon. The
calculations have been performed assuming the best candidates for the
$^{10}$Li spectrum, i.e. the spectra (c) and (d) in the upper part of
fig~\ref{fig6}.  The computed results for both spectra are completely
indistinguishable.  In part (a) of fig.~\ref{fig8} we show the core
momentum distribution. The thin line is the pure calculation while the
thick line is the result obtained after convoluting with the
experimental beam profile \cite{hum95}. In part (b) we show the
neutron momentum distribution, for which experimental data are not
available yet. In part (c) we show the angular distribution, where
$\theta$ is the angle between the $^{10}$Li momentum and the relative
momentum between neutron and core after the fragmentation. The curve
is very sensitive to the $s$ and $p$-wave mixing and the agreement is
therefore a strong indication of a $p$-wave content of about 40\%.
For both the distributions in (a) and (c) the agreement with the
experimental data is very good.

Finally we want to discuss the radius of $^{11}$Li. It has been
extracted by model computations and fitting of measured interaction
cross sections. The results depend somewhat on the assumed density
distributions and the reaction models. We obtain interaction cross
sections almost constant between 1030 and 1045 mb for a beam energy
varying between 300 and 700 MeV/nucleon on a carbon target. This
should be compared to the experimental values of $1055 \pm 14$~mb at
800 MeV/nucleon and $959 \pm 21$~mb at 400 MeV/nucleon. Also the
measured interaction cross sections for heavier targets like copper
and lead agree quite well with our model computations using the new
neutron- $^{9}$Li interaction.  Our resulting root mean square radius
is about 3.2~fm.  It should perhaps be emphasized that a larger range
and in particular another choice of radial shape of the three-body
interaction could be designed to increase the resulting radius by
fractions of a fm possibly without changing anything else.

\section{Summary and Conclusions}
\label{sec5}

The recent measurements of the properties of $^{10}$Li confine the
neutron-$^{9}$Li interaction which is decisive for the elaborate
three-body computations of $^{11}$Li. To use the new information the
$^{11}$Li model must be realistic and in particular the hyperfine
structure arising from spin-spin couplings must be included when the
measured energy spectrum of $^{10}$Li is used for comparison.

We first sketch our method of adiabatic hyperspherical expansion of
the Faddeev components to determine the structure of $^{11}$Li from a
given set of two-body interactions. Second we sketch the
participant-spectator reaction model used for all the fragmentation
cross sections. 

The neutron-neutron interaction is well known and not crucial as long
as the scattering length and effective ranges are correct. The same
could be said about the neutron-$^{9}$Li interaction but in this case
we do not have sufficient information. We believe that compelling
experimental evidence show that an $s$-state is present at around
50~keV and a $p$-state around 0.5~MeV. We proceed by constructing an
$s$-state neutron-$^{9}$Li gaussian potential of range $b$ with a
deeply bound state and one very low-lying virtual state. We also
construct a $p_{3/2}$ gaussian potential with the same range $b$ with
a bound state at -4.1~MeV. These two potentials produce a density
distribution of $^{9}$Li with the measured root mean square radius of
2.32~fm when $b \approx 2$~fm. The core density and potential radius
is now defined.

The three-body computation for $^{11}$Li with these interactions
supplemented with an appropriate spin-orbit term would lead to a
disaster since the additional two neutrons would occupy the low-lying
Pauli forbidden states of the core-neutrons. Therefore we introduce
the phase equivalent potentials for both $s_{1/2}$ and $p_{3/2}$
states. These potentials have precisely the same low-energy scattering
properties but without the Pauli forbidden bound states. For weakly
bound three-body systems this has proven to be an accurate procedure.
Although technically difficult this is possible and above all this is
a physically appropriate prescription to account for the most
important many-body effect in the present context. The radial shapes
of the potentials are not important provided the low-energy scattering
properties are correct. Therefore we are content with using
gaussians. The result is that we have two free strength parameters,
one for $s_{1/2}$ and one for $p_{1/2}$-states.

The last part of the neutron-$^{9}$Li interaction arises from the
$^{9}$Li-spin of $3/2$ coupling to the neutron angular momenta of
$s_{1/2}$ and $p_{1/2}$ resulting in two sets of spin-split states of
opposite parity but the same spins of 1 and 2. Again we maintain the
same range parameter $b$ of a gaussian potential leaving us with two
new strength parameters. These four parameters may be understood as
related to the four different $^{10}$Li-states. In addition the
properties of $^{11}$Li can now be used to constrain the parameters.

Unfortunately we must also use the three-body potential introduced to
fine tune the three-body energy which is necessary even when using a
set of two-body interactions reproducing all low-energy scattering
properties. This must be a short range potential and we make it as
neutral as possible under that assumption, i.e. we only use it to
adjust the binding energy to the measured value and not to upset other
unrelated  properties.

We know that fragmentation reaction cross sections rather strongly
select a neutron-$^{9}$Li relative $p$-wave content within $^{11}$Li
of about 40\%. We also know that most of these cross sections as well
as the $^{11}$Li binding energy and size are essentially independent
of the spin-splitting because the Pauli principle for the outer
neutrons requires at most one identical particle in each of these
spin-split states. If one is occupied the other is as well. Then the
average is the most important quantity. Thus a given $p$-wave content
selects a one dimensional relation between the average energies of the
$s_{1/2}$ and $p_{1/2}$ states. If the three-body potential had been
known or determined from independent sources both the statistically
averaged energies could have been found from $p$-wave content and
three-body binding energy. Still an educated guess using experience
provide a reasonable estimate.

We continue with spin-splitting both $s$ and $p$-waves and demanding
that one $s$-state must be at 50~keV and one $p$-state at 0.54~MeV.
With the average positions fixed this provides the spectrum of the
four $^{10}$Li states except that the ordering is yet undetermined and
the inaccuracy from the three-body potential is also still present.
Therefore we compute sensitive fragmentation data like the invariant
mass spectrum which should be able to distinguish between the various
$^{10}$Li spin assignments.  Unfortunately the precision of this
experimental data is not sufficient to exclude more than two of the
four different, but essentially discrete, spin structures.

In conclusion, the strong connections between the properties of
$^{10}$Li and $^{11}$Li are used to produce a consistent description.
Experimental constraints from both nuclei are essential. We maintain
the previous level of accuracy in the description of both structure
and breakup reactions of $^{11}$Li. The spectrum of $^{10}$Li is
confined to have average energies of $s$ and $p$-states at $0.40 \pm
0.05$~MeV. The individual $p$-states should appear at about 0.54~MeV
(most likely the 2$^+$--resonance) and at $0.35 \pm 0.15$~MeV
(probably the 1$^+$--resonance).  The $s$-states are at about 50~keV
and at $0.80 \pm 0.30$~MeV, where the present knowledge is
insufficient or too inaccurate to establish the ordering of the
1$^-$/2$^-$ doublet.  Furthermore, the three-body potential energy,
accounting for three-body polarization, off-shell effects,
core-excited states beyond two-body phenomenology or other cluster
configurations, must contribute by about $0.25 \pm 0.10$~MeV to the
binding energy.

\paragraph*{Acknowledgement.} We thank K. Riisager for continuous 
discussions and suggestions.

%\end{multicols}


\begin{thebibliography}{99}

\bibitem{zhu93} M.V. Zhukov, B.V. Danilin, D.V. Fedorov, J.M. Bang,
I.J. Thompson and J.S. Vaagen, Phys. Rep. {\bf 231} (1993) 151.

\bibitem{nie01} E. Nielsen, D.V. Fedorov, A.S. Jensen and E. Garrido, 
Phys. Rep. {\bf 347} (2001) 373.

\bibitem{wil75} K.H. Wilcox, R.B. Weisenmiller, N.A. Jelley, D. Ashery
and J. Cerny, Phys. Lett. B {\bf 59} (1975) 142.

\bibitem{ame90} A.I. Amelin {\it et al.}, Sov. J. Nucl. Phys. {\bf 52}
(1990) 782.

\bibitem{boh93} H.G. Bohlen {\it et al.}, Z. Phys. A {\bf 344}
(1993) 381.

\bibitem{tal60} I. Talmi and I. Unna, Phys. Rev. Lett. {\bf 4} (1960) 469.


\bibitem{bar77} F.C. Barker and G.T. Hickey, J. Phys. G: Nucl. Part. Phys.
{\bf 3} (1977) L23.

\bibitem{kry93} R.A. Kryger {\it et al.}, Phys. Rev. C {\bf 47} (1993) R2439.

\bibitem{you94} B.M. Young {\it et al.}, Phys. Rev. C {\bf 49} (1994) 279.

\bibitem{abr95} S.N. Abramovich, B. Ya Guzhovskii and L.M. Lazarev,
Phys. Part. Nucl. {\bf 26} (1995) 423.

\bibitem{zin95} M. Zinser {\it et al.}, Phys. Rev. Lett. {\bf 75} (1995) 1719.

\bibitem{boh97} H.G. Bohlen, W. von Oertzen, Th. Stolla, R. Kalpakchieva,
B. Gebauer, M. Wilpert, Th. Wilpert, A.N. Ostrowski, S.M. Grimes and 
M.N. Massey, Nucl. Phys. A {\bf 616} (1997) 254c.

\bibitem{tho99} M. Thoennessen {\it et al.}, Phys. Rev. C {\bf 59} (1999) 111.
               
\bibitem{cag99} J.A. Caggiano, D. Bazin, W. Benenson, B. Davids, 
B.M. Sherrill, M. Steiner, J. Yurkon, A.F. Zeller and B. Blank,
Phys. Rev. C {\bf 60} (1999) 064322.

\bibitem{cha01} M. Chartier et al., Phys. Lett. B {\bf 510} (2001) 24.

\bibitem{sel88} F. Ajzenberg--Selove, Nucl. Phys. A {\bf 490} (1988) 1.

\bibitem{aud95} G. Audi and A.H. Wapstra, Nucl. Phys. A {\bf 595} (1995) 409.

\bibitem{tos90} Y. Tosaka and Y. Suzuki, Nucl. Phys. A {\bf 512} (1990) 46.

\bibitem{joh90} L.~Johannsen, A.S.~Jensen and P.G.~Hansen,
Phys.~Lett. B {\bf 244} (1990) 357.


\bibitem{tho94} I.J. Thompson and M.V. Zhukov, Phys. Rev. C {\bf 49}
(1994) 1904.

\bibitem{wur96} J. Wurzer and H.M. Hofmann, Z. Phys. A {\bf 354}
(1996) 135.

\bibitem{des97} P. Descouvemont, Nucl. Phys. A {\bf 626} (1997) 647.

\bibitem{nun96} F.M. Nunes, I.J. Thompson and R.C. Johnson, Nucl. Phys. A 
{\bf 596} (1996) 171.

\bibitem{vin96} N. Vinh Mau and J.C. Pacheco, Nucl. Phys. A {\bf 607}
(1996) 163.

\bibitem{fed94} D.V. Fedorov, A.S. Jensen and K. Riisager, Phys. Rev. C 
{\bf 50} (1994) 2372.

\bibitem{gar96} E. Garrido, D.V. Fedorov and A.S. Jensen, Phys. Rev.
C {\bf 53} (1996) 3159.

\bibitem{gar97} E. Garrido, D.V. Fedorov and A.S. Jensen, Phys. Rev.
C {\bf 55} (1997) 1327.

\bibitem{gar98} E. Garrido, D.V. Fedorov and A.S. Jensen, Phys. Rev.
C {\bf 58} (1998) R2654.

\bibitem{sim99} H. Simon {\it et al.}, Phys. Rev. Lett. {\bf 83}
(1999) 496.

\bibitem{gar99} E. Garrido, D.V. Fedorov and A.S. Jensen, Phys. Rev.
C {\bf 59} (1999) 1272.

\bibitem{gar01} E. Garrido, D.V. Fedorov and A.S. Jensen, Nucl. Phys.
A, in press.

\bibitem{you93} B.M. Young {\it et al.}, Phys. Rev. Lett. {\bf 71}
(1993) 4124.

\bibitem{tan92} I. Tanihata {\it at al.}, Phys. Lett. B {\bf 287} (1992) 307.

\bibitem{alk96} J.S. Al-Khalili, J.A. Tostevin and I.J. Thompson,
Phys. Rev. C {\bf 54} (1996) 1843.

\bibitem{ege01} P. Egelhof, ENAM2001, Proc. Int. Conf. on Exotic
Nuclei and Atomic Masses, to be published.

\bibitem{gar97b} E. Garrido, D.V. Fedorov and A.S. Jensen, 
Nucl. Phys. A {\bf 617} (1997) 153.

\bibitem{bay87} D. Baye, J. Phys. A {\bf 20} (1987) 5529.

\bibitem{coo95} F. Cooper, A. Khare and U. Sukhatme, 
Phys. Rep. {\bf 251} (1995) 267.

\bibitem{gar99b} E. Garrido, D.V. Fedorov and A.S. Jensen, 
Nucl. Phys. A {\bf 650} (1999) 247.

\bibitem{car98} J. Carlson and R. Schiavilla, Rev. Mod. Phys. {\bf 70}
(1998) 743.

\bibitem{fed96}  D.~V.~Fedorov and A.~S.~Jensen,
Phys.~Lett. B {\bf 389} (1996) 631.

\bibitem{jen97}  A.S. Jensen, E. Garrido and D.V. Fedorov, 
Few-Body Systems, {\bf 22} (1997) 193.

\bibitem{zin97} M. Zinser {\it et al.}, Nucl. Phys. A {\bf 619} (1997) 151.

\bibitem{sim98} H. Simon, Ph.D. Thesis, Technische Universit\"{a}t, Darmstadt, 1998.

\bibitem{hum95} F. Humbert {\it et al.}, Phys. Lett. B {\bf 347} (1995) 198.

\bibitem{gei} H. Geissel and W. Schwab, private communication.

\end{thebibliography}
\end{document}